\documentclass[prl,aps,twocolumn,showpacs,floatfix]{revtex4-1}
\usepackage{amsfonts}
\usepackage{amssymb}
\usepackage{hyperref}
\usepackage{graphicx}
\usepackage{dcolumn}
\usepackage{bm,amsmath,verbatim}
\usepackage{mathrsfs}
\usepackage{color}

\usepackage[T1]{fontenc}
\usepackage[latin9]{inputenc}
\usepackage{booktabs}

\hypersetup{colorlinks,
linkcolor=blue,          citecolor=blue,        filecolor=blue,      urlcolor=blue           }

\begin{document}
\bibliographystyle{Science}
\title{Observation of heat scaling across a first-order quantum phase transition in a spinor condensate}

\author{H.-Y. Liang}
\thanks{These authors contributed equally to this work.}
\author{L.-Y. Qiu}
\thanks{These authors contributed equally to this work.}
\author{Y.-B. Yang}
\author{H.-X. Yang}
\author{T. Tian}
\author{Y. Xu}
\email{yongxuphy@tsinghua.edu.cn}
\author{L.-M. Duan}
\email{lmduan@tsinghua.edu.cn}
\affiliation{
Center for Quantum Information, IIIS, Tsinghua University, Beijing 100084,
PR China}


\begin{abstract}
Heat generated as a result of the breakdown of an adiabatic process is one of the central concepts of thermodynamics. In isolated systems, the heat can be defined as an energy increase due to transitions between distinct energy levels. Across a second-order quantum phase transition (QPT), the heat is predicted theoretically to exhibit a power-law scaling, but it is a significant challenge for an experimental observation. In addition, it remains elusive whether a power-law scaling of heat can exist for a first-order QPT. Here we experimentally observe a power-law scaling of heat in a spinor condensate when a system is linearly driven from a polar phase to an antiferromagnetic phase across a first-order QPT. We experimentally evaluate the heat generated during two non-equilibrium processes by probing the atom number on a hyperfine energy level. The experimentally measured scaling exponents agree well with our numerical simulation results. Our work therefore opens a new avenue to experimentally and theoretically exploring the properties of heat in non-equilibrium dynamics.
\end{abstract}

\maketitle

In quantum mechanics, at zero
temperature, when we drive an isolated system by tuning a system parameter,
if the driving rate is so slow such that the process is adiabatic,
the transition between energy levels cannot occur, and heat cannot be created.
Yet, when the system undergoes
a second-order QPT during the process, the relaxation time diverges and thus
adiabaticity cannot be maintained. As a result, transition between energy levels
does occur, producing the heat~\cite{Polkovnikov2008PRL,Polkovnikov2008NP}. In fact, across the transition point, the physics can be
described by the quantum Kibble-Zurek mechanism (KZM) and universal scaling laws for various
quantities, such as the temporal onset of excitations, the density of defects and the heat, are predicted~\cite{polkovnikov2011colloquium,de2010adiabatic}.
While important
aspects of the quantum KZM have been experimentally observed~\cite{chen2011quantum,Chapman2016PRL,clark2016universal,
zhang2017defect,Nature2019Lukin,Qiueaba7292}, the experimental measurement of the heat
still remains a significant challenge.

Such non-equilibrium dynamics is of crucial importance ranging from cosmology to condensed matter~\cite{kibble1980some,zurek1985cosmological,PhysRevLett.95.035701,zurek2005dynamics,PhysRevB.72.161201}.
Yet the existence of scaling laws is not limited to non-equilibrium dynamics across
a second-order QPT. It has been predicted that the scaling of some quantities can also occur
across a first-order QPT where multiple phases coexist~\cite{panagopoulos2015off,Zhong2017PRE,coulamy2017dynamics,pelissetto2017dynamic,Shimizu_2018}. In particular,
very recently, the KZM has been generalized to account for a power-law scaling
of the temporal onset of spin excitations present in a spinor condensate across
the first-order QPT~\cite{Qiueaba7292}.
The generalized KZM (GKZM) has also been experimentally observed in a spinor condensate~\cite{Qiueaba7292}.
Similar to a second-order QPT, it is natural to ask whether the heat can
still exhibit a power-law scaling across the first-order QPT.

A spinor Bose-Einstein condensate (BEC), described by a vector order parameter, provides a controllable platform to
explore non-equilibrium dynamics, and various interesting relevant phenomena have been experimentally
observed~\cite{sadler2006spontaneous,PhysRevLett.103.250401,PhysRevLett.108.035301,
PhysRevLett.105.090402,PhysRevA.84.063625,parker2013direct,Yang2019PRA,Tian2020PRL}. In some parameter regime for the condensate,
the spin and spatial degrees of freedom are decoupled
because all spin states have the same spatial wave function
under the single-mode approximation~\cite{Ueda2012}. As a consequence, the physics is significantly simplified so that
the spin degrees of freedom can be separately studied.
For an antiferromagnetic (AFM) sodium condensate, there is
a first-order QPT between a polar phase with atoms all occupying the $m_F=0$ level
and an AFM phase with atoms equally occupying the $m_F=\pm 1$ levels,
where $m_F$ is the magnetic quantum number. The system thus provides an ideal platform to study
the non-equilibrium physics across a first-order QPT.

Here we theoretically and experimentally demonstrate the existence of a power-law scaling of the heat in a sodium spinor
condensate for two dynamical processes: a one-way process
where a system is driven
from a polar phase to an AFM phase and a cyclic process where a system ends up at
the initial polar phase. For the one-way process, the power-law scaling is well characterized
by the GKZM. In experiments, we prepare an initial condensate in the polar phase and then slowly vary the quadratic Zeeman
energy $q$ by controlling magnetic and microwave fields to realize the two non-equilibrium processes.
Since the energy gap vanishes at the transition point, adiabaticity cannot be maintained no matter how
the system is driven, leading to the appearance of excitations as well as heat, which can be used as a measure of how strongly
adiabaticity is broken. Based on Refs.~\cite{Polkovnikov2008PRL,Polkovnikov2008NP}, the heat density
can be defined as an energy increase per atom
relative to the ground state at the final quadratic Zeeman energy $q_f$ over the entire process.
In experiments, it can be evaluated by measuring the atom number occupying
the $m_F=0$ level owing to a simple approximate relation between the heat density and the particle number
when $|q_f|$ is large.

\begin{figure*}
	\includegraphics[width=6in]{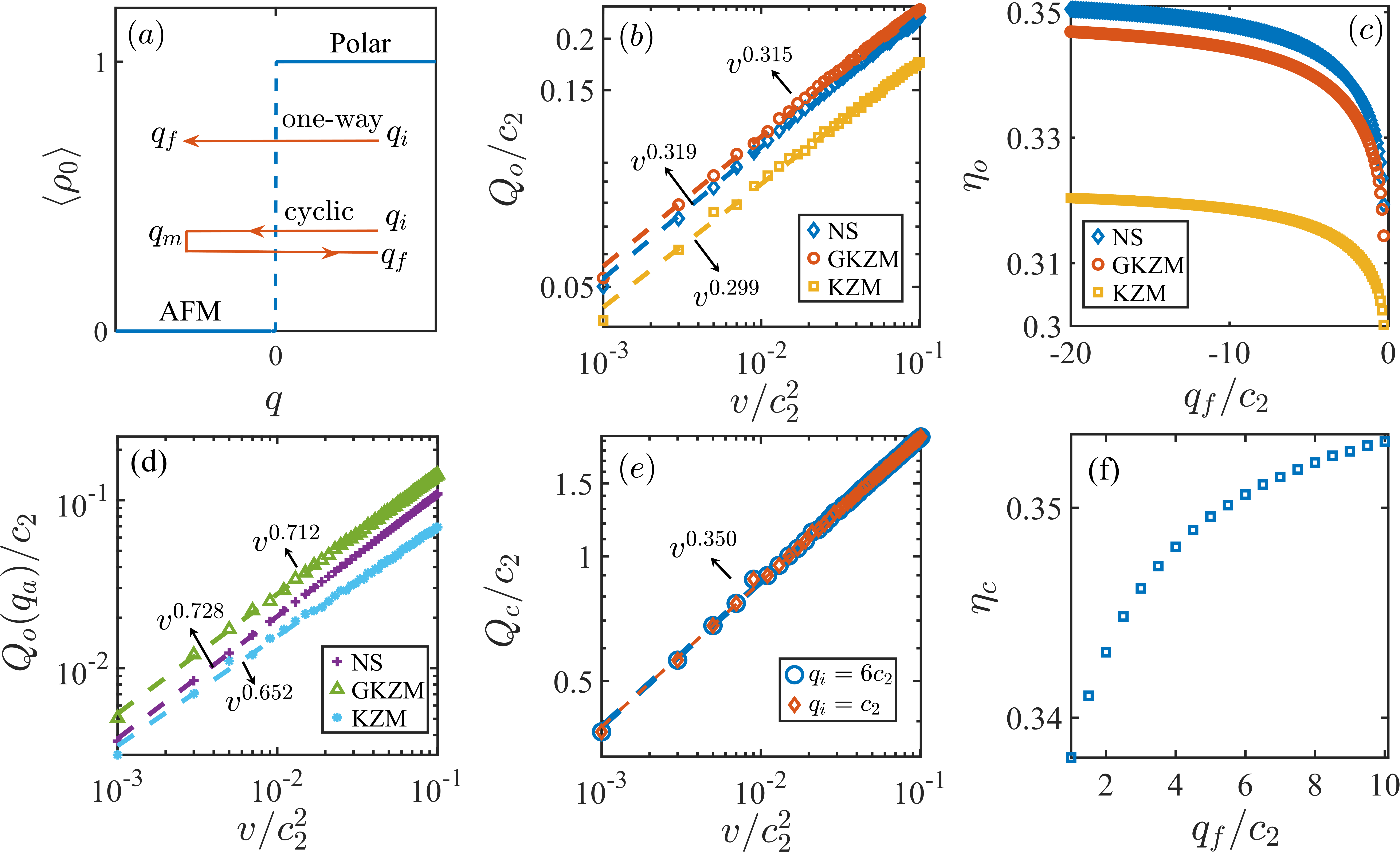}
	\caption{Schematic illustration of the quench protocol and theoretical demonstration of the heat scaling with respect to the quench rate. (a) The phase diagram depicted by $\langle \rho_0\rangle $ versus the quadratic Zeeman energy $q$.
Two linear quench protocols are considered: a one-way protocol where $q$ is varied from $q_i > 0$ to $q_f<0$
and a cyclic protocol where $q$ is changed from $q_i>0$ to $q_m<0$ and then back to $q_f>0$.
(b) The scaling of the heat density $Q_o(q_f)$ with the quench rate $v$ for a one-way process with $q_i=c_2$ and $q_f=-0.3c_2$.
(c) The scaling exponents $\eta_o$ of $Q_o(q_f)$ when $q_f$ is set to a range of different values.
(d) The scaling of the heat density $Q_o(q_a)$ with the quench rate $v$ for a one-way process with $q_i=c_2$.
Here $q_a=-vt_a$ is the corresponding quadratic Zeeman energy at the critical time $t_a$ (see the discussion in the text).
In (b-d),
the blue diamonds (purple crosses), the red circles (green triangles) and the yellow squares (cyan asterisks) represent the theoretical results for $Q_o(q_f)$ [$Q_o(q_a)$] obtained by the
numerical simulation, the GKZM and the KZM, respectively.
(e) The scaling of the heat density $Q_c(q_f)$ versus the quench rate $v$ for a cyclic process
with $q_f=6 c_2$ and $q_m=-2.5 c_2$, where blue circles and red diamonds correspond to $q_i=6c_2$  and $q_i=c_2$, respectively.
(f) The scaling exponents of the heat density for different $q_f$ for a cyclic process with $q_i=c_2$ and $q_m=-2.5 c_2$.
Here $N=1.0\times 10^4$.}
	\label{f11}
\end{figure*}

We start by considering the following Hamiltonian describing a spin-1 BEC under single-mode approximation
\begin{equation}
\label{Eq1}
\hat{H} = c_2\frac{\hat{\bf L}^2}{2N}+\sum_{m_F=-1}^{1}(qm_F^2-pm_F)\hat{a}^\dagger_{m_F}\hat{a}_{m_F},
\end{equation}
where $\hat{a}_{m_F}^\dagger$ ($\hat{a}_{m_F}$) is the Boson creation (annihilation) operator for an atom
in the hyperfine level $|F=1, m_F\rangle$,
$\hat{\bf L}$ is the total spin operator with $\hat{L}_\mu = \sum_{i, j}\hat{a}^\dagger_i(f_\mu)_{ij}\hat{a}_j$ ($\mu = x, y, z$)
and $f_\mu$ being the spin-1 angular momentum matrix, $c_2$ is the spin-dependent interaction ($c_2 > 0$ for sodium
atoms) and $N$ is the total number of atoms. Here
$p$ and $q$ describe linear and quadratic Zeeman energies, respectively.
In our experiments,
we initialize our condensates in the polar phase so that the dynamics is restricted to
the eigenspace of $L_z=0$, since the magnetization along $z$ is conserved, i.e., $[\hat{L}_z, \hat{H}] = 0$.
The linear Zeeman term $p$ thus becomes irrelevant in the dynamics even though its
value is not equal to zero in the experiment.
So the ground states of the spinor condensate correspond to the polar and AFM phases when $q>0$
and $q<0$, respectively. The phase diagram is shown
in Fig.~\ref{f11}(a) where the mean value
$\langle\rho_0\rangle$ with $\rho_0 = \hat{a}_0^\dagger \hat{a}_0/N$ taken as an order parameter
drops to zero from one at $q=0$,
indicating the occurrence of the first-order QPT.

\begin{figure}
	\includegraphics[width=3.4in]{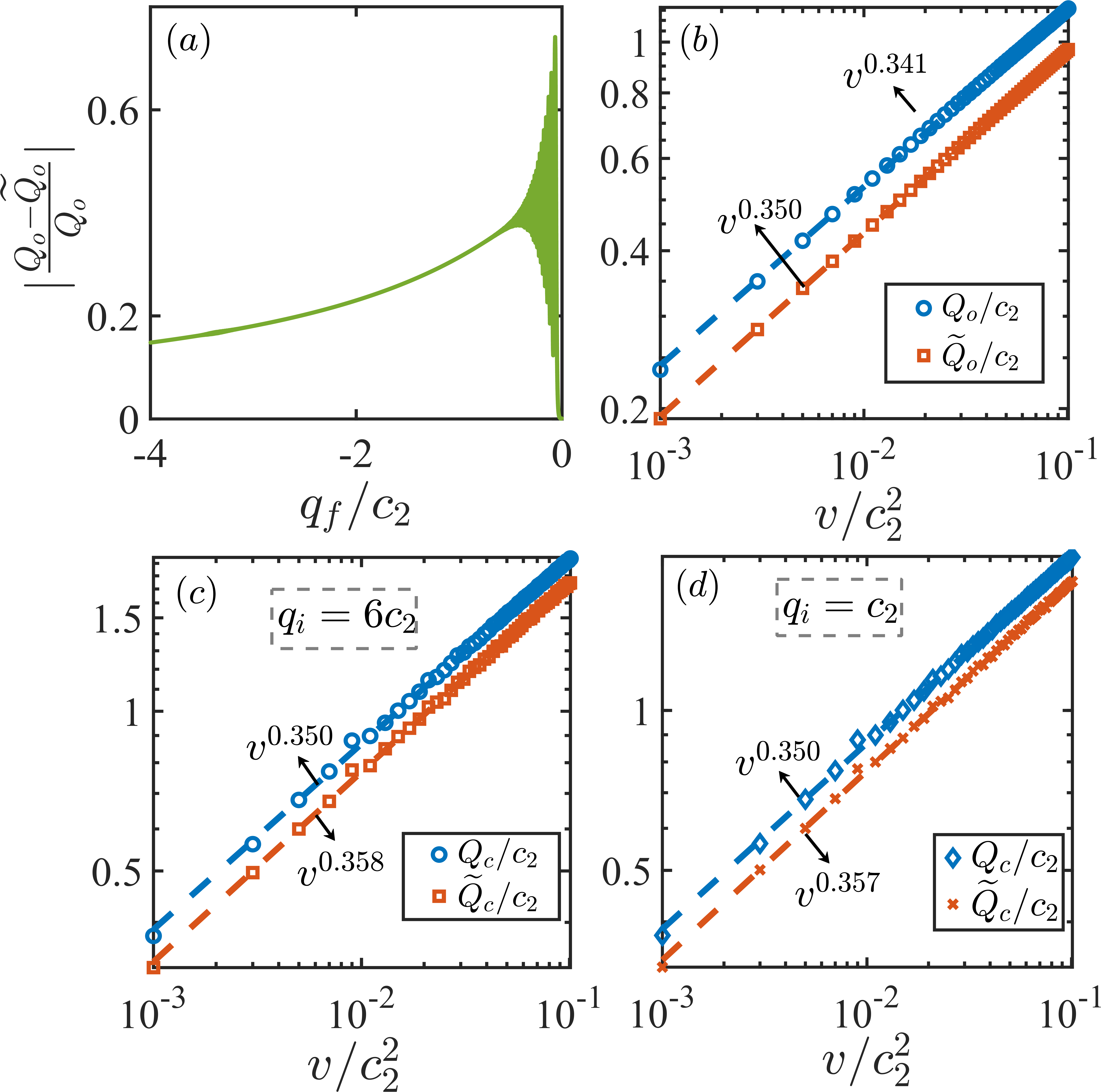}
	\caption{A comparison between the scaling of the heat density and the quasi-heat density.
(a) The relative difference of the quasi-heat density $\widetilde{Q}_o(q_f)$ compared to the heat density $Q_o(q_f)$
when $|q_f|$ increases for a fixed quench rate $v = 0.02 c_2^2$.
A comparison between the scaling of the heat density $Q(q_f)$ and the quasi-heat density $\widetilde{Q}(q_f)$
for (b) a one-way process and (c-d) cyclic processes.
In (b), $q_i=c_2$ and $q_f = -3 c_2$, in (c), $q_i=q_f=6 c_2$ and
$q_m=-2.5 c_2$, and in (d), $q_i=c_2$, $q_f=6 c_2$ and $q_m=-2.5 c_2$.
Here $N=1.0\times 10^4$. }
	\label{f12}
\end{figure}

We investigate the heat production in two types of quench processes:
a one-way process for linearly ramping $q$ from $q_i$ ($q_i>0$) to $q_f$ ($q_f<0$)
and a cyclic process for linearly ramping $q$ from $q_i$ to $q_m$ ($q_m<0$) and then back to $q_f$,
forming a cyclic process when $q_f=q_i$ [see Fig.~\ref{f11}(a)].
In both scenarios, we calculate the energy increase at the end of the quench for different
ramp rates $v$.
To numerically determine the energy of a spinor condensate,
we diagonalize the Hamiltonian under the Fock state basis in the subspace of zero magnetization,
$\{|{N}/{2},0,{N}/{2}\rangle,|{N}/{2}-1,2,{N}/{2}-1\rangle,...,|0,N,0\rangle\}$,
yielding instantaneous eigenstates $|\phi_n(q)\rangle$ $(n = 1,2, ..., {N}/{2} + 1)$ of $\hat{H}(q)$ satisfying $\hat{H}(q)|\phi_n(q)\rangle = E_n(q)|\phi_n(q)\rangle$.
We also solve
the Schr\"odinger equation $i\hbar\partial|\Psi(t)\rangle/\partial t = \hat{H}(t)|\Psi(t)\rangle$
(we take $h=1$ as a natural unit) to determine the evolving state of the spinor condensate.
The energy per atom at the end of the quench is given by
$\langle\Psi(q_f)|\hat{H}(q_f)|\Psi(q_f)\rangle=\langle H_{I}\rangle/N +q_f -q_f \langle \rho_0 \rangle$
with $H_I=c_2\hat{\bf L}^2/(2N)$ characterizing the interactions. Since the corresponding ground state energy per atom
in the AFM phase is $q_f$, the produced heat per atom over the one-way process is given by
$Q_o=\langle H_{I}\rangle/N-q_f \langle \rho_0 \rangle$. For the cyclic process, the produced
heat per atom is $Q_c=\langle H_{I}\rangle/N+q_f(1-\langle \rho_0 \rangle)$ since the ground state
energy of the polar phase is zero.

\begin{figure*}
	\includegraphics[width=\textwidth]{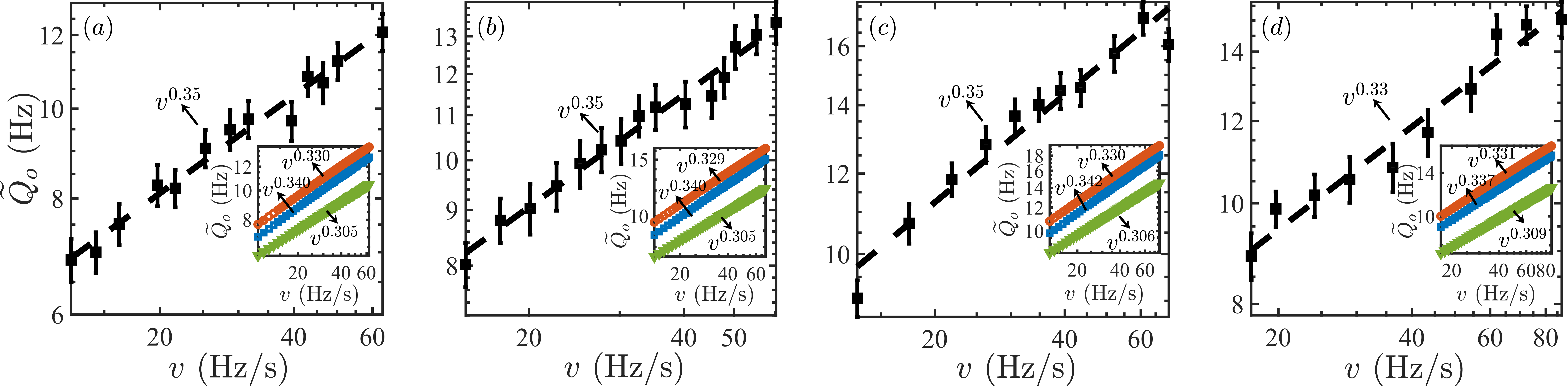}
	\caption{Experimentally observed power-law scaling of the quasi-heat density $\widetilde{Q}_o$ with respect to the ramp rate $v$ for a one-way quench.
In (a-c), we consider processes with $q_i\approx 10\,\textrm{Hz}$ and
$q_f=-21.60\,\textrm{Hz}$, $-24.47\,\textrm{Hz}$ and $-28.36\,\textrm{Hz}$, respectively
(see Appendix A for the discussion of the error
arising from the calibration of $q$).
for a spinor condensate with about $1100$ atoms corresponding to $c_2=8.1 \pm 0.9\,\textrm{Hz}$ (See Appendix B for details on
how to experimentally evaluate the value of $c_2$ and its error).
In (d), $q$ is varied from $q_i=14.33\,\textrm{Hz}$ to $q_f=-29.11\,\textrm{Hz}$ for a BEC with about $3000$ atoms and $c_2 = 11.8 \pm 0.8\,\textrm{Hz}$.
The log-log plot of the experimental data are shown as black squares with error bars.
The fitting of the data by a power-law function (black dashed line) gives the exponents of $0.35\pm 0.04$ in (a),
$0.35\pm 0.03$ in (b), $0.35\pm 0.06$ in (c) and $0.33\pm 0.06$ in (d) with $95\%$ confidence interval.
Each figure has an inset showing the theoretical results of the scaling of $\widetilde{Q}_o$
obtained by the numerical simulation (blue squares), the GKZM (red circles) and the KZM (green triangles).
The GKZM with exponents of $0.330$, $0.329$, $0.330$ and $0.331$
gives a better account of the power-law scaling with exponents of $0.340$, $0.340$, $0.342$ and $0.337$ obtained by the numerical simulation than
the corresponding exponents of
$0.305$, $0.305$, $0.306$ and $0.309$ obtained by the KZM.}
	\label{f2}
\end{figure*}

In Fig.~\ref{f11}(b), we plot our numerical simulation results of the heat density $Q_o$ for a one-way process,
remarkably showing the existence of a power-law scaling, i.e., $Q_o \propto v^\eta$
with $\eta=0.319$. This power-law scaling persists even when $q_f$ is far away from the transition point,
but the exponents change as a function of $q_f$ and increase very slowly when $|q_f|$ is large [see Fig.~\ref{f11}(c)].
The exponents are independent of $q_i$ given that two distinct $q_i's$ are connected by an adiabatic process.
For a cyclic process, we also observe the power-law scaling of the heat density $Q_c$ as shown in Fig.~\ref{f11}(e).
While the scaling does not depend on $q_i$ for the same reason, Fig.~\ref{f11}(f) shows that
the scaling exponents increase slightly with increasing $q_f$ (but they
are irrelevant of $q_m$).

It is a well-known fact that the universal scaling laws across a second-order QPT
are accounted for by the quantum KZM~\cite{PhysRevLett.95.035701,zurek2005dynamics,PhysRevB.72.161201}.
Its essential basis is the existence of impulse and adiabatic regions. Specifically,
suppose at $t=0$, $q=q_c=0$ and the system is in the polar phase.
When we linearly drive the system into the AFM phase, the system cannot respond (impulse region)
until
the response time $\tau(t_a)=1/\Delta(t_a)=t_a$, where $\Delta$ is the relevant energy gap.
When $t>t_a$, adiabaticity is restored (adiabatic region).
For a second-order QPT, the relevant energy gap
refers to the gap between the ground state and the first excited state.
Based on the KZM, the heat induced by a slow quench across the critical point is shown
to exhibit a power-law dependence on the ramp rate with the scaling exponent determined by the equilibrium critical exponents~\cite{polkovnikov2011colloquium,de2010adiabatic,You2018PRA}.

For a first-order QPT, we have demonstrated the existence of impulse and adiabatic regions
when a spinor condensate is linearly driven across the transition point~\cite{Qiueaba7292}.
Yet, in stark contrast to the KZM, the relevant energy gap is the gap between the maximally occupied state (the metastable state)
and its corresponding first excited state in
the first-order case.
For example, when $q<0$, the metastable state refers to the many-body metastable polar phase~\cite{Qiueaba7292}.
We now apply the GKZM to determine the heat scaling. To be more precise,
we use the equation $|1/\Delta(t)| = |\Delta(t)/\dot{\Delta}(t)|$ to
calculate the critical time $t_a$ and the corresponding $q_a=-v t_a$.
Since the evolving state is frozen to the initial state when $t<t_a$, the heat density can be evaluated by
$Q_o(q_f) = [\sum_n P_n E_n(q_f) - E_g(q_f)]/N$,
where $P_n=|\langle \phi_n(q_a) |\Psi(q_i)\rangle|^2$ is the
probability that the initial state $|\Psi(q_i)\rangle$ occupies the eigenstate $|\phi_n(q_a)\rangle $ of $\hat{H}(q_a)$
corresponding to the eigenenergy $E_n(q_a)$.

In Fig.~\ref{f11}(b), we show the power-law scaling of the heat density $Q_o(q_f)$ with respect to $v$ calculated by the GKZM,
which agrees very
well with the numerical simulation results. Figure \ref{f11}(c) also displays their comparison of scaling exponents versus $q_f$,
showing very good agreement
with less than $1.5\%$ discrepancy.
In comparison, we further compute the heat scaling based on the KZM,
which exhibits conspicuous discrepancy especially for large $|q_f|$ as shown in Fig.~\ref{f11}(c).
For example, when $q_f=-20c_2$, the scaling exponent obtained by the KZM has about $8.6\%$
difference from the numerical simulation results, while the GKZM only exhibits about $1.0\%$ difference.
This indicates that the GKZM gives a better account of the heat scaling law at a first-order QPT.

In fact, based on the GKZM, the heat per atom produced during a quench process ending at
$q=q_a$ is given by $-q_a$, implying that the heat scaling is determined
by the scaling of $q_a$. In Fig.~\ref{f11}(d),
we display the scaling obtained by the numerical simulation, the GKZM and the KZM, demonstrating that the GKZM gives a closer prediction of the power-law exponent to the numerical simulation result than the KZM.

To experimentally probe the heat density is a formidable task due to the complexity of the spin interactions,
which is hard to measure. Fortunately, for the one-way process, when $|q_f|$ is large, the heat density $Q_o$ is dominated by the second part,
$\widetilde{Q}_o=-q_f \langle \rho_0 \rangle$, which can be experimentally evaluated by measuring
$\langle \rho_0 \rangle$ and $q_f$. We call $\widetilde{Q}_o$ the quasi-heat density to distinguish it from the
heat density $Q_o$.
Similarly, for the cyclic process, we define the corresponding quasi-heat density as $\widetilde{Q}_c=q_f(1- \langle \rho_0 \rangle)$.
For both processes, the heat density and the quasi-heat density are related by the following equation
\begin{equation}
Q_s=\widetilde{Q}_s+\langle H_{I}\rangle/N
\end{equation}
with $s=o,c$ referring to the one-way and cyclic processes, respectively.

Figure~\ref{f12}(a) shows the decline of the relative difference between the heat density $Q_o$
and the quasi-heat density $\widetilde{Q}_o$ when $|q_f|$ is increased; the relative difference decreases to less than $20\%$ when $q_f=-3c_2$
when $v=0.02c_2^2$. In fact, the scalings
determined by these two energy increases agree much better than their energy differences
even for not very large $|q_f|$, which is experimentally achievable. For instance,
when $q_f=-3c_2$, while there exists $18\%$ difference of $\widetilde{Q}_o$ compared to $Q_o$,
their scaling exponents are in excellent agreement with only less than $3\%$ discrepancy for the one-way process [see Fig.~\ref{f12}(b)].
For the cyclic process when $q_f=6c_2$, the scaling exponent difference is also smaller than $3\%$ [see Fig.~\ref{f12}(c) and (d)].
This allows us to obtain the scaling of the heat density by experimentally measuring $\langle \rho_0 \rangle$ for relatively large
$|q_f|$.

\begin{figure*}
	\includegraphics[width=\textwidth]{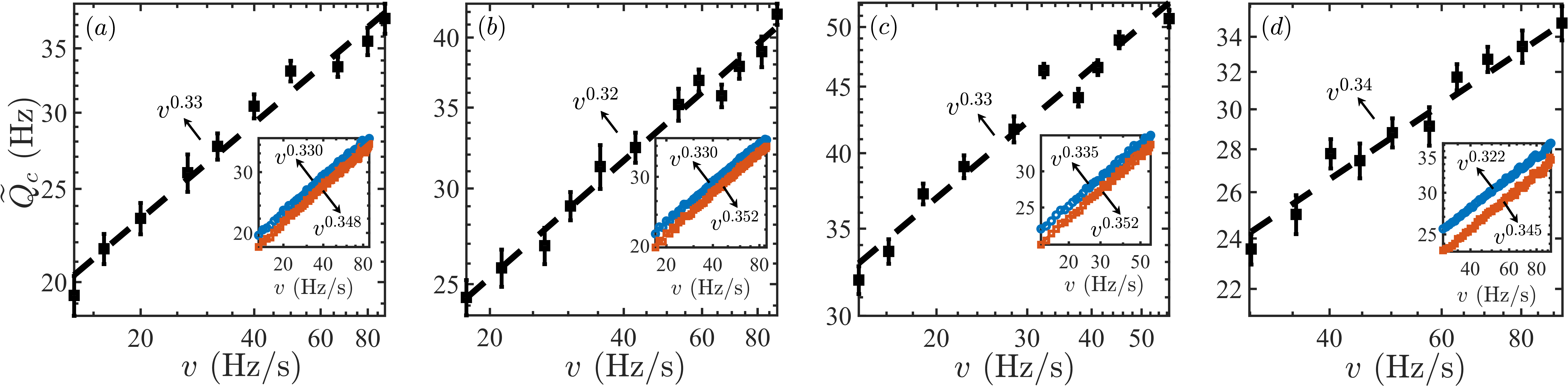}
	\caption{Experimentally observed power-law scaling of the quasi-heat density $\widetilde{Q}_c$ with respect to the quench rate for cyclic processes.
In (a-c), $q$ is linearly changed from $q_i\approx 15\,\textrm{Hz}$ to $q_m=-15.19\,\textrm{Hz}$ in (a) [$q_m=-22.00\,\textrm{Hz}$ in (b)
and $q_m=-15.49\,\textrm{Hz}$ in (c)]
and then back to the final value $q_f=60.10\,\textrm{Hz}$ [$q_f=66.45\,\textrm{Hz}$ in (c)] for a condensate with about $1100$ atoms
corresponding to $c_2=8.1 \pm 0.9\,\textrm{Hz}$. In (d), the atom number
$N=3000$ corresponding to $c_2 = 11.8 \pm 0.8\,\textrm{Hz}$, $q_i=17.55 \,\textrm{Hz}$, $q_m=-18.10 \,\textrm{Hz}$ and $q_f=66.73 \,\textrm{Hz}$.
The experimental data are plotted in the logarithmic scale, which are fitted by power-law functions,
giving the exponents of $0.33\pm 0.04$ in (a), $0.32\pm 0.03$ in (b), $0.33\pm 0.06$ in (c) and
$0.34 \pm 0.05$ in (d) under $95\%$ confidence interval. The inset of each figure shows the numerical simulation results of $Q_c$ (blue circles) and $\widetilde{Q}_c$ (red squares).}
	\label{f3}
\end{figure*}

In experiments, we produce a sodium BEC in an all-optical trap by evaporation of atoms~\cite{Yang2019PRA}.
At the evaporation cooling stage, we apply a strong magnetic field gradient
to remove the atoms on the hyperfine levels $|F=1, m_F = \pm 1\rangle$ out of the trap,
leaving all atoms on the $|F=1, m_F=0\rangle$ level.
After that, a weak and nearly resonant microwave pulse is applied to
excite the atoms from $|F=1, m_F=0\rangle$ to $|F=2, m_F=0\rangle$ (the corresponding detuning
is $\delta \simeq -6\,\textrm{kHz}$). Since the atoms on the latter level suffer a significant loss due to three body decay,
this process kicks many atoms out of the BEC cloud, resulting in less than $3000$ atoms remaining in the trap.
The reduction of the atom number allows us to avoid the unwanted relaxation to the AFM ground state when $q$ is tuned to negative values.
The atoms are then immersed in a uniform magnetic field with $q_i \approx 10 \,\textrm{Hz}$ for $2\,\textrm{s}$ to equilibrate to the polar phase.
Afterwards, we slowly decrease the magnetic field strength so as to linearly vary $q$ according to the relation
$q\approx B^2\cdot277 \,\textrm{Hz}/\textrm{G}^2$.
When $q$ is changed to around $5\,\textrm{Hz}$, we immediately switch on a microwave field with a frequency of $1.7701264\,\textrm{GHz}$ (the detuning is $\delta = -1.5 \,\textrm{MHz}$ relative to the transition from $|F=1, m_F=0\rangle$ to $|F=2, m_F=0\rangle$)
and then gradually raise its amplitude so that $q$ is linearly driven to the negative regime~\cite{YouLi2017Science} (Appendix A).
During the process, the amplitude of the microwave field is precisely controlled by a proportional-integral-derivative (PID) feedback system according to a careful calibration of $q$'s values. For each quench rate $v$, $\rho_0$ is measured
by a standard Stern-Gerlach fluorescence imaging at the end of the linear quench in each experiment
and $\langle\rho_0\rangle$ is evaluated by averaging over 40 repeated measurements (Appendix C).

In Fig.~\ref{f2}, we show the experimental results of the quasi-heat density $\widetilde{Q}_o$ with respect
to the ramp rate $v$ for the one-way quench with four different sets of quench parameters.
In (a-c), we drive
a BEC with about $1100$ atoms and $c_2 = 8.1\pm 0.9\,\textrm{Hz}$ to three distinct $q_f$.
The experimental data clearly demonstrate the existence of a power-law scaling for these different conditions.
We fit the data by a power-law function, i.e., $\widetilde{Q}_o \propto v^{\widetilde{\eta}}$,
giving the fitting exponents of $\widetilde{\eta}=0.35\pm 0.04$, $\widetilde{\eta}=0.35\pm 0.03$ and $\widetilde{\eta}=0.35\pm 0.06$, respectively.
The results agree well with the numerical simulation results with the power-law fitting exponents
of $0.340$, $0.340$ and $0.342$, respectively [see the insets of Fig.~\ref{f2}].
Here the numerically calculated exponents for $\widetilde{Q}_o$ exhibit about $10\%$ difference from the exponents of $Q_o$,
which is larger than the result shown in Fig.~\ref{f12}(b) due to
larger ramp rates considered here to reduce the relaxation to the ground states in experiments (Appendix D).
In Fig.~\ref{f2}(d), we further plot the experimental results for a BEC with roughly $3000$ atoms
and $c_2 = 11.8\pm 0.8 \,\textrm{Hz}$, showing the existence of a power-law scaling with the fitting exponent of
$0.33\pm 0.06$,
which is in good agreement with the numerically obtained exponent of $0.337$.

For a cyclic quench, similar to the one-way one, we initially prepare the condensate in the polar phase with the quadratic Zeeman
energy $q_i$ (e.g., $q_i \approx 15\,\textrm{Hz}$) provided by a uniform magnetic field and then linearly decrease $q$ to
roughly $5 \,\textrm{Hz}$ by decreasing the magnetic field strength. After that, we shine a microwave field to
the BEC to linearly vary $q$ from about $5\,\textrm{Hz}$ to $q_m$ ($q_m<0$, e.g.,
$q_m=-22\,\textrm{Hz}$) and then back to $5 \,\textrm{Hz}$ by controlling the field amplitude.
We then turn off the microwave field and raise the magnetic field strength until $q$ slowly rises to $q_f$ (e.g.,
$q_f=60.1 \,\textrm{Hz}$). Since the results do not depend on the value of $q_i$ when it is
sufficiently large so that the dynamics is adiabatic under
the quench rate at $q=q_i$ [see Fig.~\ref{f11}(e)], in experiments, we use $q_i$ and $q_f$ with $q_f>q_i$ for experimental convenience.
The entire ramping process is precisely controlled to be linear according to the calibration of $q$. Similarly, at the end of
each quench, the quasi-heat density $\widetilde{Q}$ is measured by probing $\langle\rho_0\rangle$ through the Stern-Gerlach fluorescence imaging.

For a cyclic quench, in Fig.~\ref{f3}, we show the experimental measurement of the quasi-heat density $\widetilde{Q}_c$ as a function of
the quench rate $v$ under different quench parameters.
For a BEC cloud with about $1100$ atoms
and $c_2=8.1 \pm 0.9\,\textrm{Hz}$,
the results shown in Fig.~\ref{f3}(a-c)
evidently illustrate a power-law scaling of the quasi-heat density with
fitting exponents of $0.33\pm 0.04$, $0.32\pm 0.03$ and $0.33\pm 0.06$, respectively. The exponents agree well with
the exponents of $0.348$, $0.352$ and $0.352$ numerically obtained for $\widetilde{Q}_c$,
which are larger than $0.330$, $0.330$ and $0.335$ (numerically calculated scaling exponents for $Q_c$)
by about $5.4\%$, $6.7\%$ and $5.1\%$,
respectively. This also shows that our experimental measurements cannot differentiate the slight difference between
$\widetilde{Q}_c$ and $Q_c$. Additionally, we raise the atom number to around $3000$
corresponding to $c_2 = 11.8 \pm 0.8\,\textrm{Hz}$ and perform the experiments under the quench
parameters of $q_i=17.55 \,\textrm{Hz}$, $q_m=-18.10 \,\textrm{Hz}$ and $q_f=66.73 \,\textrm{Hz}$.
Figure~\ref{f3}(d) reveals the existence of a power-law scaling with a fitting exponent of $0.34 \pm 0.05$, in
good agreement with the numerical result of $0.345$.

Our work demonstrates the first experimental observation of a power-law scaling of heat
with respect to a ramp rate
for non-equilibrium dynamics.
Two types of quench processes including one-way and cyclic processes are studied across a first-order QPT in a spinor condensate.
The experimentally measured scaling exponents for both non-equilibrium processes agree well with our
numerical simulation results.

\begin{acknowledgments}
 We thank Yingmei Liu, Ceren Dag, and Anjun Chu for helpful discussions. This work was supported by the Beijing Academy of Quantum Information Sciences, the National key Research and Development Program of China (2016YFA0301902), Frontier Science Center for Quantum Information of the Ministry of Education of China, and Tsinghua University Initiative Scientific Research Program. Y. Xu also acknowledges the support from the start-up fund from Tsinghua University, the National Natural Science Foundation
of China (11974201) and the National Thousand-Young-Talents Program.
\end{acknowledgments}

\begin{widetext}
\section*{Appendix A: Calibration of the quadratic Zeeman energy $q$}

\setcounter{equation}{0} \setcounter{figure}{0} \setcounter{table}{0} %
\renewcommand{\theequation}{A\arabic{equation}} \renewcommand{\thefigure}{A%
\arabic{figure}} \renewcommand{\thetable}{A%
\arabic{table}} \renewcommand{\bibnumfmt}[1]{[#1]} \renewcommand{%
\citenumfont}[1]{#1}

In the experiment, the quadratic Zeeman energy $q$ is contributed by
both the magnetic and microwave fields so that
\begin{equation}
	\label{EqS1}
	q = q_B + q_M,
\end{equation}
where $q_B$ and $q_M$ are generated by the magnetic and microwave field, respectively. Specifically,
$q_B$ is determined by the magnetic field strength $B$ through
$q_B \approx B^2 \cdot 277 \, (\textrm{Hz}/\textrm{G}^2)$.
Based on the relation $p \approx B \cdot 700 \,(\textrm{KHz/G})$ with $p$ being the linear Zeeman energy,
$B$ can be measured by probing $p$ through a Rabi oscillation between the Zeeman energy level $|F=1, m_F=0\rangle$ and $|F=2, m_F=-1\rangle$.
A microwave field at large frequency detuning shifts the energy of the Zeeman levels due to the AC Stark effect and contributes to the quadratic Zeeman energy as
\begin{equation}
	\label{EqS2}
	q_M = \frac{\Delta E_{m_F=+1} + \Delta E_{m_F=-1} - 2 \Delta E_{m_F=0}}{2}
\end{equation}
with
\begin{equation}
	\label{EqS3}
	\Delta E_{m_F} = \frac{1}{4}\sum_{k=-1,0,+1}\frac{\Omega^2_{m_F, m_F+k}}{\Delta_{m_F, m_F+k}},	
\end{equation}
where $\Omega_{m_F, m_F+k}$ is the Rabi frequency for the resonant transition from $|F=1, m_F\rangle$ to $|F=2, m_F+k\rangle$ and $\Delta_{m_F, m_F+k}$ is the microwave detuning relative to the transition between these energy levels.

In the experiment, the magnetic field strength is controlled by a voltage $V_B$ that determines the magnitude of the current flowing through the Helmholtz coils. In Fig.~\ref{fS1}(a), we show the measured $q_B$ as a function of the voltage $V_B$, which is well fitted by a parabola function, thus allowing us to linearly change $q_B$ by controlling $V_B$.

To determine the microwave quadratic Zeeman energy $q_M$, we experimentally evaluate the Rabi frequencies $\Omega_{0,-1}$, $\Omega_{0,1}$ and $\Omega_{-1,-1}$ by probing the Rabi oscillations for a resonant transition between the two corresponding energy levels. Other Rabi frequencies can be calculated according to the following relations
\begin{align}
	&\Omega_{0, 1} = \sqrt{3}\Omega_{-1,0}, \ \Omega_{1,2} = \sqrt{6}\Omega_{-1,0},\\
	&\Omega_{1, 1} = \Omega_{-1, -1} = \frac{\sqrt{3}}{2}\Omega_{0,0},\\
	&\Omega_{0,-1} = \sqrt{3}\Omega_{1,0}, \ \Omega_{-1, -2} = \sqrt{6}\Omega_{1,0}.
\end{align}
We now fix the microwave's frequency at $1.7701264\, \textrm{GHz}$ with a detuning $\Delta_{0,0} = -1500\,\textrm{kHz}$ for the transition from $|F=1, m_F=0\rangle$ to $|F=2, m_F=0\rangle$. $q_M$ is then calculated based on Eq.~(\ref{EqS2}). Since the magnetic field still exists with
$q_B \approx 5.0 \, \textrm{Hz}$ when the microwave pulse is applied,
the total quadratic Zeeman energy
$q=q_M+q_B$. Because the Rabi frequencies depend on the microwave field amplitude, we can control the amplitude
to vary $q_M$ and $q$. In our experiment, we stabilize the amplitude of the microwave pulse by a PID system and
calibrate the values of $q$ at several different microwave amplitudes controlled by the feedback voltage $V_f$.
The measured $q$ with respect to $V_f$ is displayed in Fig.~\ref{fS1}(b) with a parabola fitting to the data
allowing for a linear ramp of $q$ by tuning $V_f$.

\begin{figure}[h]
	\includegraphics[width=4.5in]{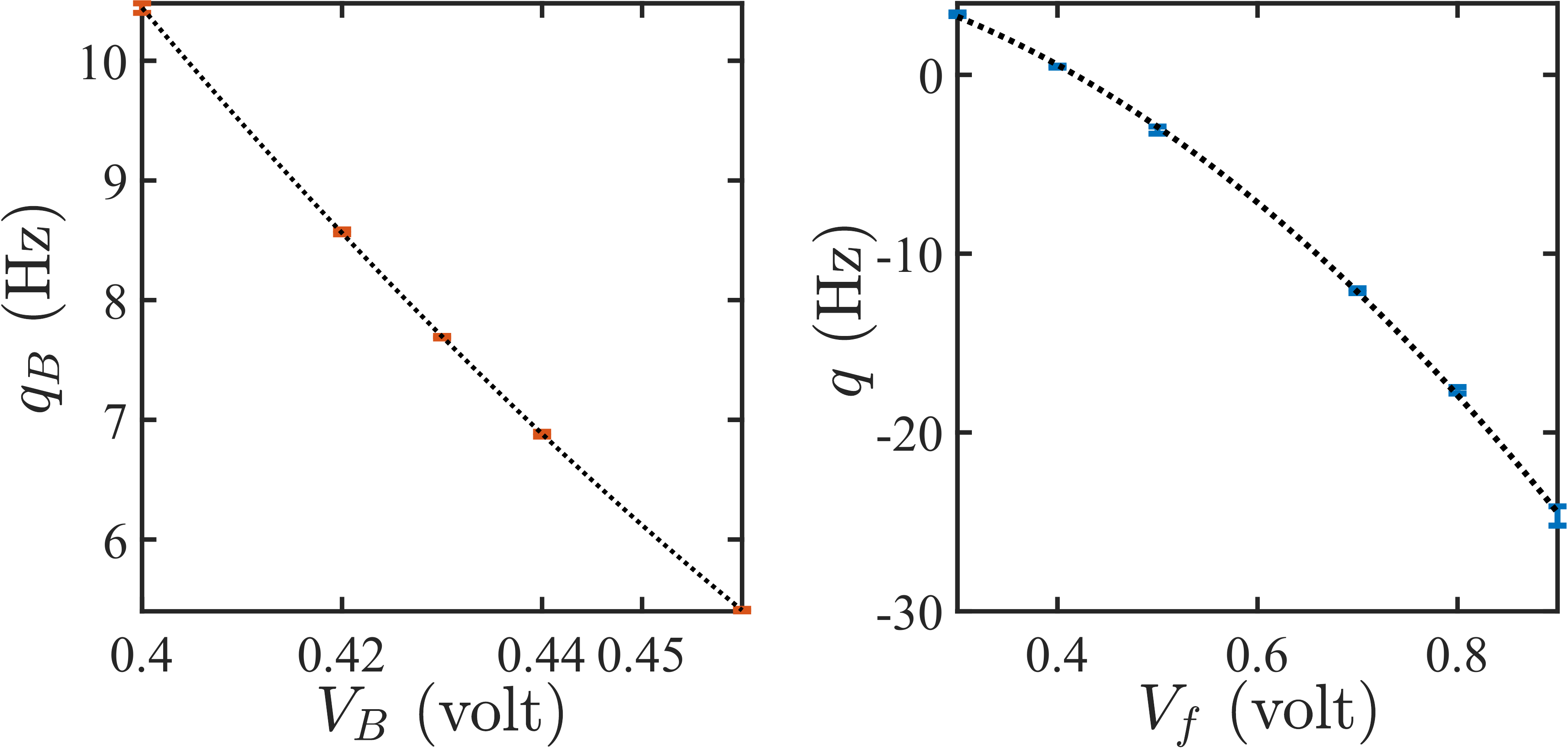}	
	\caption{(Color online) Experimental calibration of the quadratic Zeeman energy $q$. (a) Experimentally measured $q_B$ (black crosses) as a function of the voltage $V_B$. (b) Experimentally measured $q$ (black crosses) as a function of the feedback voltage $V_f$ of the PID system.
The experimental data are fitted by the red dashed parabola curves.}\label{fS1}
\end{figure}

To estimate the error between the calibrated $q$ and the true value of $q$, we apply a sudden quench method to
measure the transition point $q_c = 0$. Specifically, we initialize the condensate in the polar phase with $q_i>0$ and then suddenly change $q$ to $q_f$. After $500 \, \textrm{ms}$' evolution, we probe the atom populations on the $m_F=0$ state. If $q_f>0$, then all atoms should remain on the $m_F=0$ state; otherwise, if $q_f<0$, atoms will show up on the $m_F = \pm 1$ level. In the experiment, the sudden quench of $q$ is realized by switching on the microwave field with the frequency of $1.7701264\, \textrm{GHz}$ and a certain amplitude controlled by $V_f$. Since $q_f$ decreases as $V_f$ increases [see Fig.~\ref{fS1}(b)], we can find a maximum $V_f^{\textrm{max}}$ so that all the atoms stay on the $m_F=0$ state and a minimum $V_f^{\textrm{min}}$ so that atoms begin to show up on the $m_F = \pm 1$ states.
The calibrated value of the transition point $q_c$ is thus $q_{\textrm{cali}} = [q(V_f^{\textrm{min}})+q(V_f^{\textrm{max}})]/2$,
resulting in a calibration error of $\delta q=q_{\textrm{cali}}-q_c=q_{\textrm{cali}}$ for each set of data.
We summarize the calibration error $\delta q$ for 21 days' measurements in Table~\ref{tS1}.

\begin{table}
	\centering
	\begin{tabular}{c|c|c|c}
		\hline
		$q(V_f^{max})$(Hz) & $q(V_f^{min})$(Hz) & $q_{cali}$(Hz) & $\delta q$(Hz)\\
		\hline
		-1.03 & -1.11 & -1.07 & \\
		0.43 & 0.35 & 0.39 & \\
		0.15 & 0.07 & 0.11 & \\
		-0.62 & -0.85 & -0.74 & \\
		-0.83 & -1.06 & -0.95 & \\
		0.35 & 0.28 & 0.32 & \\
		-1.16 & -1.37 & -1.27 & \\
		0.13 & 0.06 & 0.10 & \\
		-0.12 & -0.17 & -0.15 & \\
		0.20 & 0.13 & 0.17 & \\
		-1.43 & -1.63 & -1.53 & $-0.45 \pm 0.70$\\
		-1.23 & -1.44 & -1.34 & \\
		0.40 & 0.33 & 0.37 & \\
		0.29 & 0.22 & 0.26 & \\
		-1.04 & -1.25 & -1.15 & \\
		0.31 & 0.20 & 0.26 & \\
		-1.13 & -1.34 & -1.24 & \\
		0.32 & 0.25 & 0.29 & \\
		0.01 & -0.27 & -0.13 & \\
		-1.32 & -1.53 & -1.43 & \\
		-0.64 & -0.84 & -0.74 & \\
    \hline
	\end{tabular}
	\caption{Summary of the measured values of $q(V_f^{\textrm{max}})$ and $q(V_f^{\textrm{min}})$ obtained
through sudden quench experiments performed during 21 days. The calibrated value of $q$ in each row is obtained by $q_{\textrm{cali}}=(q(V_f^{\textrm{max}})+q(V_f^{\textrm{min}}))/2$, giving the calibration error $\delta q=\bar{q}_{\textrm{cali}}+\sigma$
where $\bar{q}_{\textrm{cali}}$ is the mean value of $q_{\textrm{cali}}$ and $\sigma$ is the standard deviation.}\label{tS1}
\end{table}

\section*{Appendix B: Measurement of $c_2$}

\setcounter{equation}{0} \setcounter{figure}{0} \setcounter{table}{0} %
\renewcommand{\theequation}{B\arabic{equation}} \renewcommand{\thefigure}{B%
\arabic{figure}} \renewcommand{\thetable}{B%
\arabic{table}} \renewcommand{\bibnumfmt}[1]{[#1]} \renewcommand{%
\citenumfont}[1]{#1}

The spin-dependent interaction coefficient $c_2$ can be measured by observing the spin oscillation, i.e., time evolution of $\rho_0$ when the condensate is initially prepared to the state $|\rho_1, \rho_0, \rho_{-1}\rangle$ with $\rho_1 = \rho_{-1}$ and $0<\rho_0<1$. According to the mean-field theory, where the quantum fluctuations are neglected and the operators are replaced by their expectation values, the spin-mixing dynamical equations for the spin-1 condensate are given by~\cite{You2005PRA}
\begin{equation}
	\label{EqS4}
	\dot{\rho}_0 = \frac{2c_2}{\hbar}\rho_0\sqrt{(1-\rho_0)^2-m^2}\,\textrm{sin}\theta,
\end{equation}

\begin{equation}
	\label{EqS5}
	\dot{\theta} = -\frac{2q}{\hbar} + \frac{2c_2}{\hbar}(1-2\rho_0) + (\frac{2c_2}{\hbar})\frac{(1-\rho_0)(1-2\rho_0)-m^2}{\sqrt{(1-\rho_0)^2-m^2}}\, \textrm{cos}\theta,
\end{equation}
where $m = \rho_{1} - \rho_{-1}$ is the magnetization and $\theta = \theta_{+} + \theta_{-} - 2\theta_{0}$ is the relative phase. By fixing the quadratic Zeeman energy $q$ and $m \approx 0$, we simulate the time evolution of $\rho_0$ to find the value of $c_2$ that best fits the experimental results. $c_2$ is then obtained by averaging over 5 measurements for 5 distinct $q$.
In Fig.~\ref{fS2}(a) and (b), we show the experimental and theoretical results of the time evolution of $\rho_0$ with the initial state being
$|\rho_1, \rho_0, \rho_{-1}\rangle = |0.28, 0.46, 0.26\rangle$ for $N=1100$
and $|\rho_1, \rho_0, \rho_{-1}\rangle = |0.30, 0.44, 0.26\rangle$ for $N=3000$, respectively. In Table~\ref{tS2}, we summarize the five measured results for different atom numbers, giving the mean value of $c_2$ of $8.1\, \pm 0.9\, \textrm{Hz}$ for $N=1100$ and $11.8\, \textrm{Hz} \pm 0.8\, \textrm{Hz}$ for $N=3000$.
\begin{figure}		
	\includegraphics[width=0.48\textwidth]{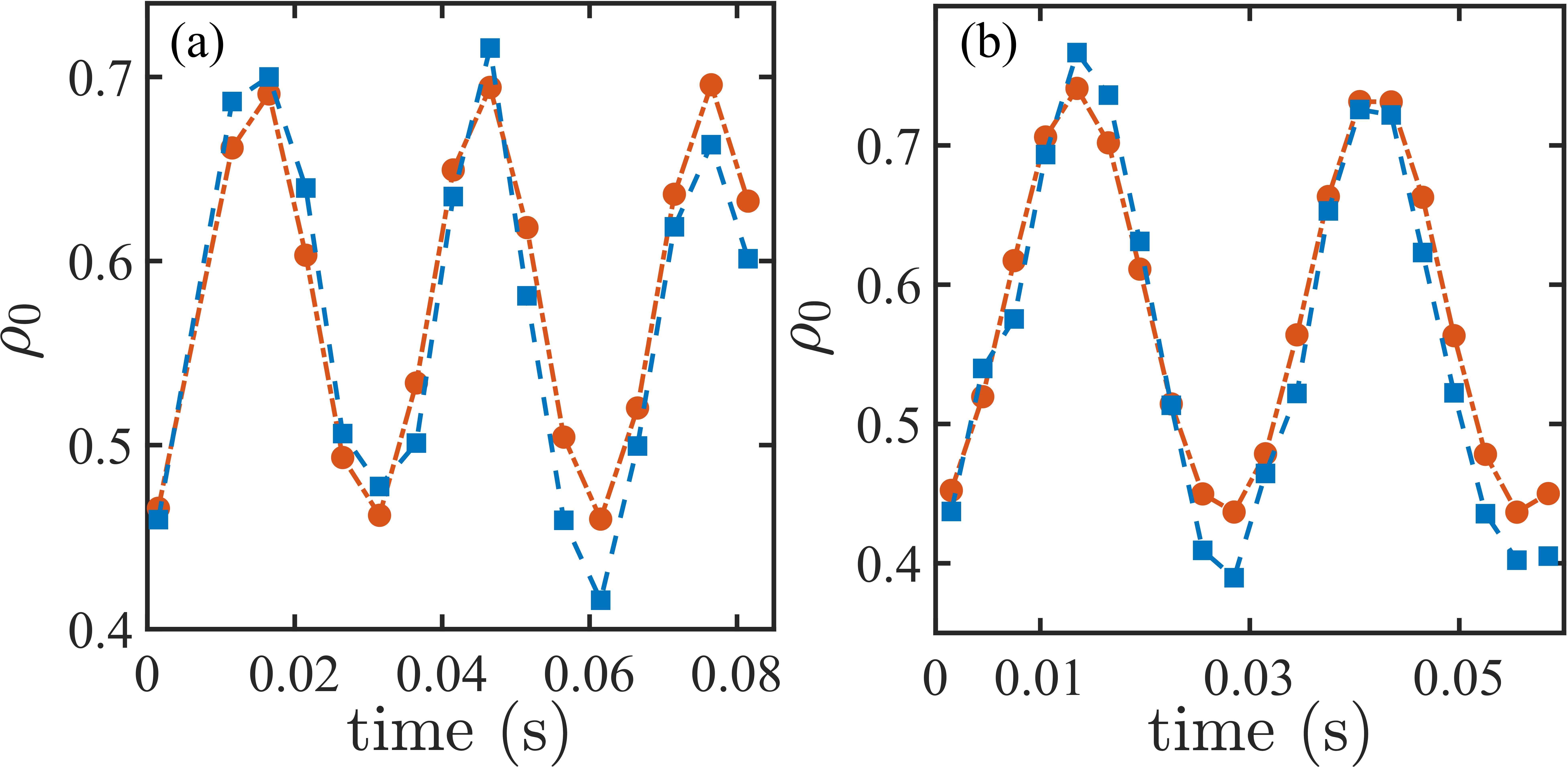}

	\caption{(Color online) Spin oscillation measurements of $c_2$ by fitting the experimentally observed time evolution of $\rho_0$ (blue squares)
by theoretical simulations (red circles). In (a) [(b)], the initial state is prepared to $|0.28, 0.46, 0.26\rangle$ ($|0.30, 0.44,0.26\rangle$) for a system with about
$1100$ ($3000$) atoms and $q=14.31\,\textrm{Hz}$ ($q=14.22\, \textrm{Hz}$).}
	\label{fS2}
\end{figure}
\begin{table}
	\centering
	\begin{tabular}{c|c|c|c|c}
		\hline
		$q(Hz)$ & $\theta (\times\pi)$ & $c_2(Hz)$ & $\bar{c}_2(Hz)$ & N \\
		\hline
		14.31 & 0.99 & 8.0 &  \\
		17.15 & 0.97 & 7.4 &  \\
		24.09 & 1.01 & 9.8 & $8.1 \pm 0.9Hz$ & 1100\\
		27.69 & 1.01 & 8.2 &  \\
		31.91 & 1.15 & 7.3 &  \\
		\hline
		9.44 & 0.95 & 12.6 & \\
		11.62 & 0.91 & 11.4 & \\
		14.22 & 0.95 & 11.4 & $11.8 \pm 0.8Hz$ & 3000 \\
		17.17 & 0.97 & 10.8 & \\
		20.45 & 1.05 & 12.9 & \\
		\hline
	\end{tabular}
	\caption{Summary of the datasets for $c_2$'s measurements.}\label{tS2}
\end{table}

\begin{figure}
	\includegraphics[width=0.48\textwidth]{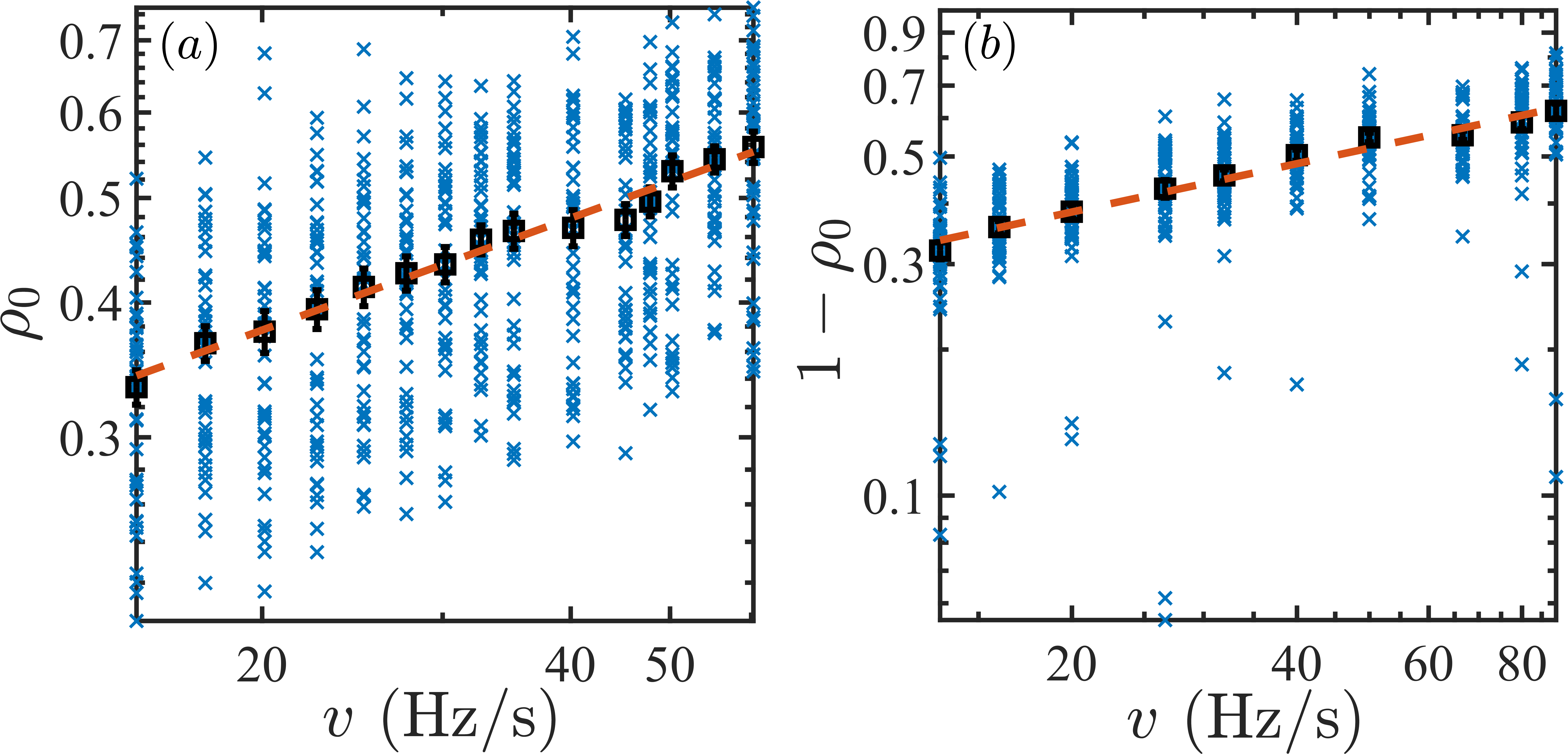}
	\caption{(Color online) Log-log plots of the measured $\rho_0$ (blue diagonal crosses) and their mean values (black squares).
(a) A one-way quench process with $N\approx1100$, $q_i = 9.96\,\textrm{Hz}$ and $q_f=-24.47\, \textrm{Hz}$.
(b) A cyclic quench process with $N\approx1100$, $q_i = 15.04\, \textrm{Hz}$, $q_m=-15.19\, \textrm{Hz}$ and $q_f=60.1\, \textrm{Hz}$.
The red dashed lines denote the linear fits of $\langle\rho_0\rangle$. }\label{fS3}
\end{figure}

\section*{Appendix C: Experimental measurement of $\rho_0$}

\setcounter{equation}{0} \setcounter{figure}{0} \setcounter{table}{0} %
\renewcommand{\theequation}{C\arabic{equation}} \renewcommand{\thefigure}{C%
\arabic{figure}} \renewcommand{\thetable}{C%
\arabic{table}} \renewcommand{\bibnumfmt}[1]{[#1]} \renewcommand{%
\citenumfont}[1]{#1}

In experiments, we measure $\rho_0$ by the standard Stern-Gerlach fluorescence imaging for different ramp rates.
Fig.~\ref{fS3}(a) shows the measured data (labelled by blue diagonal crosses) of $\rho_0$ for a one-way process where $q$ is linearly varied from $9.96\,\textrm{Hz}$ to $-24.47\,\textrm{Hz}$. In this case,
the atom number $N$ is restricted to about $1100$ corresponding to a fluorescence count in the range of $2.5\times10^9$ and $2.8\times10^9$. $\langle\rho_0\rangle$ at each ramp rate is calculated by averaging over measurements repeated 40 times, which is plotted as black squares
in the figure.
The error bars of $\langle\rho_0\rangle$ (the error in the mean) originate from the quantum fluctuations and the measurement fluctuations and is evaluated as $\sigma/\sqrt{40}$~\cite{StatisticBook}, where $\sigma$ is the standard deviation of the 40 samples. For the one-way quench process, $\widetilde{Q}_o = -q_f \, \langle\rho_0\rangle$ and thus the error bar of $\widetilde{Q}_o$ can be evaluated by
\begin{equation}
	\delta\widetilde{Q}_o = \sqrt{(\langle\rho_0\rangle^{\textrm{var}} + \langle\rho_0\rangle^2)(q_f^{\textrm{var}} + q_f^2) - (\langle\rho_0\rangle\, q_f)^2},
\end{equation}
where the superscript $\textrm{var}$ denotes the variance of a quantity. Since there is an error of $-0.45\,\textrm{Hz}$
for $q_f$, we make a correction of $+0.45\,\textrm{Hz}$ to $q_f$ with the variance $q_f^{var} = 0.70^2 \,(\textrm{Hz})^2 = 0.49\,(\textrm{Hz})^2$ according to the calibration of $q$ in Table.~\ref{tS1}.
In Fig.~\ref{fS3}(b), we also display the original data of $\langle \rho_0\rangle$ for a cyclic process, where $q$ is changed from $15.04\, \textrm{Hz}$ to $-15.19\, \textrm{Hz}$ and then back to $60.1\, \textrm{Hz}$ at different ramp rates for a system with roughly $1100$ atoms.

\section*{Appendix D: Finite quench rate effects on scaling exponents}

\setcounter{equation}{0} \setcounter{figure}{0} \setcounter{table}{0} %
\renewcommand{\theequation}{D\arabic{equation}} \renewcommand{\thefigure}{D%
\arabic{figure}} \renewcommand{\thetable}{D%
\arabic{table}} \renewcommand{\bibnumfmt}[1]{[#1]} \renewcommand{%
\citenumfont}[1]{#1}

In Fig.~\ref{fS4}, we provide the numerical simulation results of $Q_o$, showing about $10\%$ difference for $\widetilde{Q}_o$
compared with $Q_o$. This discrepancy is larger than the result shown in Fig. 2(b) in the main text.
We attribute this discrepancy to the finite ramp rates.
As shown in Table~\ref{tS3},
the relative difference is larger for a range of quench rates with larger values due to larger contribution of interactions
to total energy for a fixed $q_f$.
In our experiments, both $q_f$ and quench rate $v$ that can be taken are limited by the applied microwave
field which can induce the relaxation of the condensate to the AFM ground state when its amplitude
is strong or it is applied for a long time.
To reduce the relaxation effect, we take the minimum $q_f$ as $-29.11\,\textrm{Hz}$
and the slowest ramp rate as $10\, \textrm{Hz/s}$ corresponding to about $3\,\textrm{s}$ for an quench process
during which the microwave field is shined to vary the quadratic Zeeman energy.

\begin{figure*}[t]
	\includegraphics[width=1.0\textwidth]{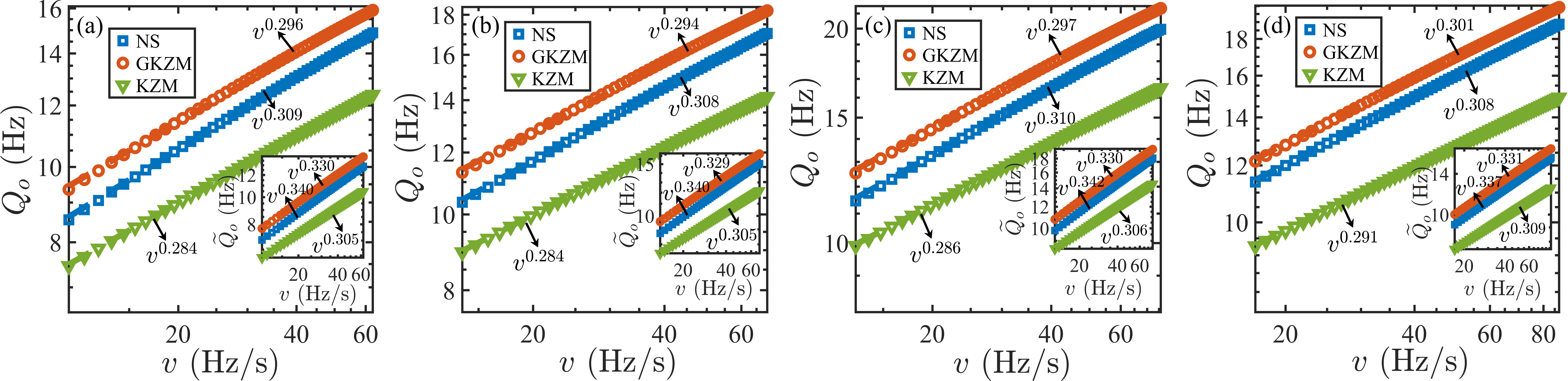}
	\caption{(Color online) Comparison of the theoretical results of $Q_o$ and $\widetilde{Q}_o$ corresponding to Fig.~3 in
the main text. The quench parameters are the same as in Fig.~3. The results by numerical simulation, the KZM and the GKZM are
plotted as blue squares, green triangles and red circles, respectively. The inset shows the theoretical results of $\widetilde{Q}_o$ as a comparison to those of $Q_o$.}
	\label{fS4}
\end{figure*}

\begin{table}
	\centering
	\begin{tabular}{c|c|c|c}
		\hline
		$v(c_2^2)$ & $\eta$ & $\widetilde{\eta}$ & $|\eta-\widetilde{\eta}|/\eta$\\
		\hline
		$0.001 \sim 0.01$ & 0.370 & 0.373 & 0.81$\%$ \\
		$0.011 \sim 0.1$ & 0.351 & 0.356 & 1.42$\%$ \\
		$0.1 \sim 1$ & 0.315 & 0.343 & 8.89$\%$ \\
		\hline
	\end{tabular}
	\caption{Finite quench rate effects on the scaling exponents for $Q_o$ and $\widetilde{Q}_o$ in the one-way quench process with $q_i=c_2$, $q_f=-3c_2$ and $N=1100$. Different ranges of ramp rates $v$ give different scaling exponents $\eta$, $\widetilde{\eta}$ and their relative difference.}\label{tS3}
\end{table}

\end{widetext}

\end{document}